\documentclass[11pt]{article}
\usepackage{epsfig}
\newcommand{\N}{{\scriptscriptstyle N}}

\newcommand{\be}{\begin{equation} }
\newcommand{\ee}{\end{equation} }
\newcommand{\bea}{\begin{eqnarray}}
\newcommand{\eea}{\end{eqnarray}}

\def\I_M{{I_{\scriptscriptstyle M\times M}}}

\def\N{{\scriptscriptstyle N}}

\def\tr{\mbox{tr}}

\def\L{{\cal L}}

\def\half{{\textstyle{\frac{1}{2}}}}

\def\N{{\scriptscriptstyle N}}
\def\g{{g^{\prime}}}
\def\o{{\scriptscriptstyle 0}}

\begin{document}
\begin{titlepage}
\title{\vskip -60pt
{\small
\begin{flushright}
hep-th/0203017
\end{flushright}}
\vskip 45pt  Description  of  identical  particles   via gauged
matrix models \\ ~{}\\~{}:\,a generalization of the
Calogero-Sutherland system} \vspace{4.0cm}
\author{\\
\\
\\Jeong-Hyuck Park\,\thanks{Electronic correspondence: jhp@kias.re.kr}}
\date{}
\maketitle \vspace{-1.0cm}
\begin{center}
\textit{Korea Institute for Advanced Study}\\ \textit{207-43 Cheongryangri-dong, Dongdaemun-gu}\\ \textit{Seoul 130-012, Korea}
\end{center}
\vspace{2.0cm}
\begin{abstract}
\noindent We elaborate the idea that  the  matrix  models equipped
with the gauge symmetry  provide a natural framework to describe
identical particles. After demonstrating the general
prescription, we study an exactly solvable harmonic oscillator
type gauged matrix model. The  model gives   a generalization of
the Calogero-Sutherland system where the strength of the inverse
square  potential is not fixed but dynamical bounded by below.
\end{abstract}

\thispagestyle{empty}
\end{titlepage}
\newpage

\setcounter{footnote}{0}

\section{Introduction}
Recent advances in the noncommutative field theory have enabled us to realize that the quantum Hall system is  closely  related to the noncommutative
Chern-Simons theories. After the pioneering work by Susskind \cite{0101029}, utilizing the fact that  the noncommutative field theories can be formulated
by matrices of infinite size, Polychronakos proposed a finite matrix Chern-Simons model  for the description of  the finite number of electrons in the
quantum Hall system \cite{0103013}. Soon after, a matrix version of the Laughlin's wavefunction \cite{Laughlin} was recovered  by Hellerman and Raamsdonk
\cite{0103179}.

In this paper, we wish to clarify the underlying  principles  of
the above  matrix model approach to the quantum Hall system in a
more general setup. We elaborate the idea  that the
 matrix models equipped with \textit{gauge symmetry} provide a natural framework to
describe identical particles (see e.g. \cite{9902157,PLB26629}).

After  demonstrating the general prescription, we explicitly
study an exactly solvable harmonic oscillator type gauged matrix
model. We show the model gives a generalization of the
Calogero-Sutherland system where the strength of  the inverse
square potential is not fixed but dynamical bounded by below.

\section{Description of Identical Particles}
One of the intrinsic properties of the fundamental particles in nature is   the very fact that they are identical.  Namely  it is in principle impossible
to identify  each individual particle at different time slices. In the ordinary quantum mechanics, the conventional way of incorporating this idea is to
anti-symmetrize the wavefunctions over the particle indices by hand.

Matrix gauge theories provide  a more natural framework to describe  identical particles.   To see this we first consider  the classical Lagrangian for
$N$ electrons
\begin{equation}
\L_{\o}=\sum_{a=1}^{N}\,T(x_{a},\dot{x}_{a})-V(x_{1},x_{2},\cdots,x_{N})\,, \label{classic}
\end{equation}
where $x\in{\bf R}^{D}$, the $D$-dimensional `space', and   the potential is totally symmetric over the particle indices.  Formally introducing a diagonal
$N\times N$ matrix, $X=\mbox{diag}(x_{1},x_{2},\cdots,x_{N})$,   one can always rewrite the Lagrangian in terms of  the matrix
\begin{equation}
\L_{\o}=\tr\!\left[T(X,\dot{X})\right]-{\cal V}\left({\tr f_{l}(X)}\right)\,,
\label{fact}
\end{equation}
where we used the fact  that the symmetric potential   can be always written in terms of the traces of   certain set of functions, ${\tr
f_{l}(X)}=\sum_{a=1}^{N}\,f_{l}(x_{a})$. We put the proof at the end of this section.   In particular, when $x_{a}$ carries no spatial index as in the
$D=1$ case,  one can simply set $f_{l}(x)=x^{l},\,1\le l\le N$.

Allowing the off-diagonal elements with the restriction, $X$ being hermitian to ensure the eigenvalues to be real,  we
encounter a new physical system. First, the action acquires a $\mbox{U}(N)$ symmetry for which  the matrices transform in
the adjoint representation
\begin{equation}
\begin{array}{cc}
X\,\rightarrow\,UXU^{\dagger}\,,~~~&~~~U\in\mbox{U}(N)\,.
\end{array}
\label{XU}
\end{equation}
We note that the permutation group relevant to the   relabeling  of the electrons is a subgroup of $\mbox{U}(N)$.    As
electrons are identical,  the labeling is not  physical so that  the permutation symmetry is an auxiliary one. This
suggests to gauge away the  $\mbox{U}(N)$ symmetry in the matrix formalism. Namely we introduce an auxiliary matrix,
$A_{\o}$ and  replace the ordinary time derivative by a covariant time derivative
\begin{equation}
D_{t}X=\dot{X}+i\,[A_{\o},X]\,.
\end{equation}
Then what $A_{\o}$ brings new is the Gauss' constraint or the equation of the motion for $A_{\o}$,  which gives  the quantum   generator of the
$\mbox{U}(N)$ symmetry from the Noether theorem. Since we embed the discrete permutation group into the continuous group, the reparameterization  of the
electrons can be now realized by quantum operators. At the quantum level the constraint is to be imposed on the wavefunctions and this will mode out the
auxiliary symmetry. We emphasize the point that the $\mbox{U}(N)$ symmetry (\ref{XU}) can be now time dependent, and physically this amounts that we are
requiring the physics to be invariant  under not only time independent but also time dependent reparameterization of the particles.  Once again, it is in
principle impossible or  meaningless to identify each individual particle at different time slices.

One  interesting ``generalization" is to add a term  into the action which is linear in $A_{\o}$ \cite{PLB26629,0108145}
\begin{equation}
\kappa_{{\rm {\scriptscriptstyle bare}}}\,\tr A_{\o}\,. \label{CS}
\end{equation}
While writing the Gauss' constraint at the quantum level there always occurs  an ordering ambiguity,   since the
constraint  contains the product of $X$ and its conjugate momentum. The ambiguity amounts to adding an identity matrix to
the Gauss' constraint up to a factor. Thus, considering the above term (\ref{CS}) is not a mere ``generalization" but
rather  a natural requirement. After all, the final form of the  Gauss' constraint which is to be imposed on the quantum
wavefunctions  should be written in the normal ordered way with the \textit{physical} coefficient, $\kappa$
\begin{equation}
\underbrace{\mbox{U$(N)$ generator}}_{\textrm{\small normal~ordered}}=\kappa\,I_{\N\times\N}\,.
\end{equation}
After all, the actual integer value of the physical coefficient,
$\kappa$ is given by the superselction rule.

The consistency at the quantum level requires the physical coefficient, $\kappa$ to be an integer
\cite{PLB26629,0108145}.  At the quantum level, the Gauss' constraint generates unitary transformations,
$U=e^{i\Lambda},\,\Lambda^{\dagger}=\Lambda$ on all the arguments in the wavefunction \cite{0101029}
\begin{equation}
|\Psi^{\prime}\rangle=e^{i\kappa\,{\tr}\Lambda}\,|\Psi\rangle\,. \label{GW}
\end{equation}
Taking the particular choice, $\Lambda=\mbox{diag}(2\pi,0,0,\cdots,0)$  gives  the identity matrix, $U=1_{\N\times\N}$,
and   apparently the Gauss' constraint on the wavefunctions works successfully only  for integer, $\kappa$. Essentially
this quantization is identical to that of  the coefficient in the noncommutative Chern-Simons theories \cite{0102188}.

Nevertheless, non-zero $\kappa$ is  yet problematic for the finite matrix models. As the matrices are  in the adjoint representation,  the central U($1$)
transformation would leave the wavefunction invariant, and  this is clearly inconsistent with Eq.(\ref{GW})  for non-zero $\kappa$. Curing the problem
requires the presence of  new variables in other representations. A natural candidate is a complex bosonic vector, $\phi$ in the fundamental
representation so that $D_{t}\phi=\dot{\phi}+iA_{\o}\phi$. As we will see later,  this new variable governs the strength of the intrinsically existing
repulsive potential in the matrix gauge theories.

Introducing  the complex vector,  the gauged matrix model can be  self-consistent and is now of the general form
\begin{equation}
\L=\tr\!\left[T(X,D_{t}{X})+\kappa_{{\rm {\scriptscriptstyle bare}}}\,A_{\o}\right]-{\cal V}\left({\tr f_{a}(X)}\right)+ \mbox{vector~parts}\,.
\end{equation}
As worked out in \cite{0103179,0108145} the general quantum wavefunction satisfying the Gauss' constraint consists  of
$\mbox{U}(N)$ invariant part times $|\kappa|$ products of $\mbox{SU}(N)$ invariant parts (see Eq.(\ref{wave})). Due to
the latter, the wavefunction is  an eigenstate of the particle exchange operator of the eigenvalue, $(-1)^{\kappa}$
capturing the identical nature of the particles.

The  prescription to give the physical meaning to any expectation value in the matrix model is the map
\begin{equation}
\begin{array}{ccc}
\langle\Psi|\,\tr{f(X)}\,|\Psi\rangle~~&\longleftrightarrow&~~\displaystyle{\int}{\rm d}x^{{\scriptscriptstyle
D}}~\rho(x)f(x)\,,
\end{array}
\label{map}
\end{equation}
where $\rho(x)$ is the corresponding physical density function,
while $f(x)$ is an arbitrary function.   As we turn on the
off-diagonal components in $X$ and introduce the vectors, the
present non-matrix system is not simply equal to the initial one
(\ref{classic}). Physically, the off-diagonal element,
$X^{a}{}_{b},\,a\neq b$ corresponds to the interaction between
the two particles, $a$, $b$, and integrating out the off-diagonal
components generates an inverse square type repulsive potential.
Thus the  matrix gauge theories intrinsically contain an
repulsive potential among the particles.   However, this can be
in principle eliminated in the matrix model by adding the counter
term written in the matrix form, if necessary.

It is worth to note that due to the Gauss' constraint on the wavefunction (\ref{GW})  any matrix valued expectation must
be $\mbox{U}(N)$ invariant, and hence
\begin{equation}
\langle\Psi|\,f(X)\,|\Psi\rangle=\displaystyle{\frac{1}{N}}\langle\Psi|\,\tr{f(X)}\,|\Psi\rangle\,I_{\N\times\N}\,.
\end{equation}

For the systems of  kinetic terms quadratic in time derivatives in the $D$ dimensional space  the phase space for the matrices has   the dimension,
$2DN^{2}$ so that subtracting the Gauss' constraint and the  gauge symmetry, the physical degrees of freedom  for the matrices is $2(D-1)N^{2}+1$. On the
other hand the vector has the linear in $N$ degree of freedom. Hence when $D=1$  the total degrees of freedom is linear in $N$,  otherwise it is
quadratic in $N$. Nevertheless if we include the BFSS type potential, $\tr [X^{I},X^{J}]^{2}$ \cite{BFSS}, at the low energy   limit the matrices tend to
commute each other making the ``off-diagonal components'' negligible. This will result in linear degrees of freedom at low energy. In fact, in a special
low energy limit the particles form a collinear motion \cite{0108145,9705047}.  On the other hand, for the kinetic term linear in time derivatives like
the $D=2$ matrix Chern-Simons model, the dimension of the phase space is halved and the total degrees of freedom in linear in $N$. \newline

\textit{The proof of the fact}\\
Consider a symmetric function depending on $N$ particles' coordinates, $\vec{x}_{a}, a=1,2,\cdots, N$  of the form
\begin{equation}
F(x_{1},\cdots,x_{\N})=\displaystyle{\sum_{p}}\,f_{1}(x_{p_{1}})f_{2}(x_{p_{2}})\cdots f_{\N}(x_{p_{\N}})\,,
\end{equation}
where $f_{1},\cdots,f_{\N}$ are functions of one particle coordinate and the sum is over the $N!$  permutations so that the function is apparently
symmetric over the particle indices.  General symmetric functions can be written in terms of this kind of  symmetric functions. For example, the Coulomb
interaction can be written as a fraction of such two functions. In the below we show that $F$ can be written in terms of $\tr\,f_{l}(X)$ with
$X=\mbox{diag}(x_{1},x_{2},\cdots,x_{\N})$. We prove this by the mathematical induction on the number of non-constant functions in
$\{f_{1},f_{2},\cdots,f_{\N}\}$ which we denote by $\#_{f}$. If $\#_{f}=1$, it is easy to see
\begin{equation}
F(x_{1},\cdots,x_{\N})=(N-1)!\,\tr\,f_{\alpha}(X)\displaystyle{\prod_{\beta\neq\alpha}}\,f_{\beta}\,,
\end{equation}
where $f_{\alpha}$ is the only one non-constant function. Hence the statement holds for $\#_{f}=1$. Now we assume that
$F$ can be written in terms of $\tr\,f_{l}(X)$ for $\#_{f}<n$ cases, and   consider $F$ in the $\#_{f}=n$ case. We let
with out loss of generality $f_{1},\cdots,f_{n}$ be the $n$ non-constant functions and set
\begin{equation}
F^{\prime}\equiv
F-(N-n)!\,\left(\displaystyle{\prod_{\alpha=1}^{n}}\,\tr\,f_{\alpha}(X)\right)\left(\displaystyle{\prod_{\beta >
n}}\,f_{\beta}\right)\,.
\end{equation}
It is crucial to note that $F^{\prime}$ belongs to the classes, $\#_{f}<n$ so that $F^{\prime}$ and hence $F$ can be written in terms of $\tr\,f_{l}(X)$.
This completes our proof.

\section{Generalization of Calogero-Sutherland System}
The $D=1$ gauged matrix model we consider here is   the harmonic oscillator type. With the column vector, $\phi$, and  the row vector,
$\bar{\phi}\equiv\phi^{\dagger}$, the Lagrangian is
\begin{equation}
\L=\tr\!\left[\half m(D_{t}X)^{2}-\half gX^{2}+m D_{t}\phi D_{t}\bar{\phi}  -\g\phi\bar{\phi}+\kappa_{{\rm {\scriptscriptstyle bare}}}\,A_{\o}\right]\,.
\label{L}
\end{equation}
If $g=\g$, writing a ${(N+1)\times(N+1)}$ matrix ${(\!{\tiny\begin{array}{c} X\,\phi\\
\bar{\phi}\,\,\!\,0\end{array}}\!)}$,  the model can be regarded as the truncation of the bigger matrix model with the
broken gauge symmetry, ${(\!{\tiny\begin{array}{c} A\,\,0\\ 0\,\,\!\,0\end{array}}\!)}$,
$\mbox{U}(N+1)\rightarrow\mbox{U}(N)$.

With the conjugate momenta, $P=mD_{t}X,$ $p_{\phi}=mD_{t}\phi$ we define the following dimensionless quantities
\begin{equation}
\begin{array}{ll}
C=\frac{1}{\sqrt{2}}\left((mg)^{\frac{1}{4}}X+i(mg)^{-\frac{1}{4}}P\right)\,,~~~&~~~ \eta_{\pm}=\frac{1}{\sqrt{2}} \left((m\g)^{\frac{1}{4}}\phi\pm
i(m\g)^{-\frac{1}{4}}p_{\phi}\right)\,.
\end{array}
\end{equation}
The standard quantization shows the non-vanishing commutators are
\begin{equation}
\begin{array}{ccc}
{[C^{a}{}_{b},\bar{C}^{c}{}_{d}]=\delta^{a}{}_{d}\delta_{b}{}^{c}\,,}~~&~~
[\eta^{a}{}_{+},\bar{\eta}_{+b}]=\delta^{a}{}_{b}\,,~~&~~[\eta^{a}{}_{-},\bar{\eta}_{-b}]=-\delta^{a}{}_{b}\,.
\end{array}
\label{quantization}
\end{equation}
Thus, $\bar{C},\bar{\eta}_{+},\eta_{-}$ and
$C,\eta_{+},\bar{\eta}_{-}$   are respectively creation and
annihilation operators (our  somewhat unconventional notation for
$\eta_{-}$ is    to keep the consistent $\mbox{U}(N)$ index
notation.). The appearance of the two sets of vector-valued
harmonic oscillators is due to the fact that the kinetic term for
the vector is written with the  quadratic time derivatives. By
introducing one more vector field, one could equivalently rewrite
the vector parts in the action to contain two  kinetic terms
linear in time derivatives having the opposite signs (see
\cite{Morariu:2001qa} for further generalization of the  kinetic
terms linear in time derivatives).

In terms of the operators, the Gauss' constraint reads in the
normal ordered form,
\begin{equation}
\bar{C}^{c}{}_{b}C^{a}{}_{c}-\bar{C}^{a}{}_{c}C^{c}{}_{b}+\bar{\eta}_{+b}\,\eta^{a}_{+}-\eta_{-}^{a}\,\bar{\eta}_{-b}
=\kappa\,\delta^{a}{}_{b}\,, \label{Gauss}
\end{equation}
where the left hand side generates the $\mbox{U}(N)$
transformations, while $\kappa$ on the right hand side is now of
the physical value. The Hamiltonian is
\begin{equation}
\begin{array}{ll}
H&=\tr\!\left[\half\sqrt{\frac{g}{m}}(C\bar{C}+\bar{C}C)
+\sqrt{\frac{\g}{m}}(\eta_{+}\bar{\eta}_{+}+\eta_{-}\bar{\eta}_{-})\right]
\\ {}&{}\\
{}&=\sqrt{\frac{g}{m}}{\tr(\bar{C}C)} +\sqrt{\frac{\g}{m}}\left(\bar{\eta}_{+}\eta_{+}
+\tr(\eta_{-}\bar{\eta}_{-})\right)+\half\sqrt{\frac{g}{m}}\,N^{2}+\sqrt{\frac{\g}{m}}\,N\,.
\end{array}
\label{Hamiltonian}
\end{equation}
Here the second expression is written in the normal ordered fashion so that $\tr(\bar{C}C),$  $\bar{\eta}_{+}\eta_{+},$   $\tr(\eta_{-}\bar{\eta}_{-})$
are  the number operators and $\frac{1}{2}\sqrt{\frac{g}{m}}\,N^{2}+\sqrt{\frac{\g}{m}}\,N$ is the zero-point  fluctuation of the energy.\newline

The exact wavefunctions satisfying the $\mbox{U}(N)$ covariance condition (\ref{GW})  due to the Gauss' constraint are
for $\kappa\geq 0$ and $\kappa<0$ respectively \cite{0103179,0108145}
\begin{equation}
|\Psi\rangle=\left\{
\begin{array}{l}
G\!\times\!\left(\epsilon^{a_{1}a_{2}\cdots a_{\N}}\bar{\eta}_{+a_{1}}(\bar{\eta}_{+}\bar{C})_{a_{2}}
\cdots(\bar{\eta}_{+}\bar{C}^{\N-1})_{a_{N}}\right)^{\kappa}|0\rangle\\  {}\\
G\!\times\!\left(\epsilon_{a_{1}a_{2}\cdots a_{\N}}\eta_{-}^{a_{1}}(\bar{C}\eta_{-})^{a_{2}}
\cdots(\bar{C}^{\N-1}\eta_{-})^{a_{N}}\right)^{-\kappa}|0\rangle,
\end{array}
\right. \label{wave}
\end{equation}
where  $G=G(\tr\bar{C}^{l},\bar{\eta}_{+}\bar{C}^{m}\eta_{-})$ is
an arbitrary function of the   $\mbox{U}(N)$ invariant building
blocks which are made of the creation operators only,
$\tr\bar{C}^{l}$ and $\bar{\eta}_{+}\bar{C}^{m}\eta_{-}$ with
$1\leq l\leq N$, $0\leq m$.   The trivial $G$ corresponds to the
ground state whose energy is
\begin{equation}
E_{0}=\sqrt{\frac{\g}{m}}(|\kappa|+1)N+\frac{1}{2}\sqrt{\frac{g}{m}}((|\kappa|+1)N^{2}-|\kappa|N)\,.
\end{equation}

In particular when $\kappa=0$, the vacuum, $|0\rangle$ is the ground state, and  in this case we can  calculate the exact density function, $\rho_{0}(x)$
in Eq.(\ref{map}). First, using the large $N$ behaviour,
\begin{equation}
\langle 0|\tr(C+\bar{C})^{2m}|0\rangle\sim \frac{(2m)!}{m!(m+1)!}N^{m+1}\,,
\end{equation}
one can check that the Fourier mode for $\rho_{0}(x)$ is given by the Bessel function, $J_{1}$. Realizing the Fourier transformation of  a half circle is
$J_{1}$ we get for large $N$
\begin{equation}
\rho_{0}(x)=\frac{1}{\pi}\sqrt{2N}(mg)^{\frac{1}{4}}\,\mbox{Re}\!
\left(\sqrt{1-\textstyle{\frac{(mg)^{\frac{1}{2}}}{2N}}x^{2}}\right)\,.
\end{equation}
In fact, this density function is identical to that in the
Calogero-Sutherland system \cite{Calogero-Sutherland}.  An
intuitive way to see this result is to note the close relation
to    the matrix quantum Hall system with the confining harmonic
potential \cite{0103013}, where the density function is constant
on a disc. Essentially our density function is the one-dimensional
projection of it. For general $\kappa$ at the center, $x=0$ we
expect
\begin{equation}
\displaystyle{\lim_{N\rightarrow\infty}\,\frac{\left[\rho_{\kappa}(0)\right]^{2}}{2N}
=\frac{\sqrt{mg}}{\,\pi^{2}(|\kappa|+1)}}\,,
\end{equation}
and  this is an analogue of the fractional filling factor, $\nu=1/(|\kappa|+1)$  in the quantum Hall system.  The ``+1'' in the denominator is again due
to the zero-point fluctuation or the Vandermonde determinant.

Henceforth we discuss the classical dynamics of the system focusing on the $\kappa\ge 0$ case.  The equations of motion are
\begin{eqnarray}
&0=mD_{t}D_{t}X+gX\,,\label{EoMX}\\ {}\nonumber\\
&0=mD_{t}D_{t}\phi+\g\phi\,\,\left(\Leftrightarrow\, 0=D_{t}\eta_{\pm}\pm
i\textstyle{\sqrt{\frac{\g}{m}}}\,\eta_{\pm}\right)\,.\label{EoMphi}
\end{eqnarray}

As in \cite{0108145} we choose the gauge such that $X$ is diagonal and $\eta_{+}^{a}$ is real and non-negative.  For the
simplicity of notation we define a matrix,  $K^{a}{}_{b}\equiv\eta_{+}^{a}\bar{\eta}_{+b}-\eta_{-}^{a}\bar{\eta}_{-b}\,$.
Now the Gauss' constraint determines $\eta_{+}^{a}$ and the off-diagonal components of $A_{\o}$ completely
\begin{equation}
\begin{array}{cc}
\eta_{+}^{a}=\sqrt{\kappa+|\eta^{a}_{-}|^{2}}\,,~~~&~~~
 A_{\o}^{a}{}_{b}=\displaystyle{\frac{K^{a}{}_{b}/m}{(x_{a}-x_{b})^{2}}}~~~~~\mbox{for~}a\neq b\,.
\label{offA}
\end{array}
\end{equation}
Substituting these into the Hamiltonian we obtain a generalized
Calogero-Sutherland model (cf.\cite{cf})
\begin{equation}
H_{{\rm {\scriptscriptstyle eff}}}=\displaystyle{\sum_{a=1}^{N}}\,\textstyle{\frac{1}{2m}}p_{a}^{2}+\half
gx^{2}_{a}+2\textstyle{\sqrt{\frac{\g}{m}}}\bar{\eta}_{-a}\eta^{a}_{-}\,
+\,\displaystyle{\sum_{a>b}}\,\displaystyle{\frac{|K^{a}{}_{b}|^{2}/m}{(x_{a}-x_{b})^{2}}}\,, \label{Heff}
\end{equation}
where $K^{a}{}_{b}$ is now a function of $\eta_{-}$ only
\begin{equation}
K^{a}{}_{b}=\sqrt{(\kappa+|\eta^{a}_{-}|^{2})(\kappa+|\eta^{b}_{-}|^{2})}-\eta_{-}^{a}\bar{\eta}_{-b}\,. \label{Keta}
\end{equation}

 Apparently  the complex vector, $\eta_{-}$ gives the novelty. Namely  the strength of the repulsive potential, $|K^{a}{}_{b}|$ between two
particles is not fixed but varies. The Schwarz inequality shows $|K^{a}{}_{b}|\geq\kappa$ and the saturation occurs when $\eta_{-}^{a}$ is  independent
of the particle index, $a$. In particular, for the finite energy configurations with the large $\g$ limit,   $\eta_{-}$  vanishes classically and  the
saturation occurs.

The corresponding Lagrangian is with (\ref{Keta})
\begin{equation}
\label{Leff}
\begin{array}{ll}
\L_{{\rm {\scriptscriptstyle eff}}}=&\displaystyle{\sum_{a=1}^{N}}\,\textstyle{\frac{1}{2}}m\,\dot{x}_{a}^{2}-\half
gx^{2}_{a}-i\bar{\eta}_{-a}\dot{\eta}_{-}^{a}-2\textstyle{\sqrt{\frac{\g}{m}}}\bar{\eta}_{-a}\eta^{a}_{-}
-\displaystyle{\sum_{a>b}}\,\displaystyle{\frac{|K^{a}{}_{b}|^{2}/m}{(x_{a}-x_{b})^{2}}}\,.
\end{array}
\end{equation}
The minus sign for the kinetic term of  $\eta_{-}$ is essentially
from  Eq.(\ref{quantization}).   Consistency requires that the
dynamics of this generalized Calogero-Sutherland model must agree
with the full equations of motion of the  matrix model
(\ref{EoMX},\,\ref{EoMphi}). Some straightforward manipulation
with Eq.(\ref{offA}) can show that  only when the diagonal
component of $A_{\o}$ is given by
\begin{equation}
A_{\o}^{a}{}_{a}=-\textstyle{\sqrt{\frac{\g}{m}}}-\displaystyle{\sum_{b\neq a}}\, \half
(A^{a}_{\o}{}_{b}+A^{b}_{\o}{}_{a})(\eta^{b}_{+}/\eta_{+}^{a})\,,
\end{equation}
we get
$D_{t}\eta_{+}+i\textstyle{\sqrt{\frac{\g}{m}}}\,\eta_{+}=0$, and
the matrix equations  of motion  reduce to those of the present
generalized Calogero-Sutherland model
\begin{equation}
\begin{array}{l}
0=m\ddot{x}_{a}+gx_{a}-\displaystyle{\sum_{b\neq a}\,\frac{2|K^{a}{}_{b}|^{2}/m}{(x_{a}-x_{b})^{3}}}\,, \\ {}\\
0=i\dot{\eta}_{-}^{a}+2\textstyle{\sqrt{\frac{\g}{m}}}\,\eta_{-}^{a} +\displaystyle{\sum_{b\neq
a}\,\frac{\textstyle{\frac{\partial~~}{\partial\bar{\eta}_{-a}}}|K^{a}{}_{b}|^{2}/m}{(x_{a}-x_{b})^{2}}}\,.
\end{array}
\end{equation}
The former comes from the diagonal components in (\ref{EoMX}) while the off-diagonal components  vanish  due to
(\ref{EoMphi}), $\left(mD_{t}D_{t}X+gX\right)^{a}{}_{b} =-i(D_{t}K)^{a}{}_{b}/(x_{a}-x_{b})=0$,  $a\neq b$.

Especially   when the effective charge saturates, $|K^{a}{}_{b}|=\kappa$, we get the solution for the two-particle system in a closed form
\begin{equation}
\eta^{a}_{-}=c\,e^{2i\sqrt{\frac{\g}{m}}\,(t+t_{\o})}\,,
\end{equation}
and $x_{1},\,x_{2}$ are
\begin{equation}
l_{{\scriptscriptstyle CM}}\sin(\textstyle{\sqrt{\frac{g}{m}}}\,t)\,\pm
l\sqrt{\sin(\textstyle{2\sqrt{\frac{g}{m}}}\,t+\theta)+\sqrt{1+{ k^{2}/(4gml^{4})}}}\,.
\end{equation}

It would be interesting to see any quantum correction to the inverse square potential in Eq.(\ref{Leff}).   When the  vector freezes it should reduce to
$\kappa^{2}\rightarrow \kappa(\kappa+1)$ \cite{0107168}. In any case, since $\kappa$ is  an integer, the charge of the potential is quantized.

\section{Conclusion}
We have argued that the gauged matrix  models provide a natural
framework to describe  identical particles since the particle
indices  which are not physical are gauged. The models
intrinsically contain the interaction among the identical
particles. Especially for $D=1$ or the $[X^{I},X^{J}]\rightarrow
0$ limit, it corresponds to the inverse square type  repulsive
potential.  As an example we have considered an exactly solvable
harmonic oscillator type gauged matrix model. We have obtained
all the exact  quantum wavefunctions and demonstrated that  the
model is a generalization of the Calogero-Sutherland system where
the strength or the charge of the  potential is not fixed but
dynamical bounded by below.   Furthermore, when the saturation
occurs the charge  is quantized.

The model we considered is non-relativistic as the kinetic term
is quadratic in time derivatives, and this is essentially the
reason  why we get the inverse square potential. It is very much
desirable to find a  matrix gauge model where the intrinsic
interaction is given by the inverse potential or the Coulomb
interaction.   In this sense,  it is worth  to study the
relativistic matrix  model and analyze its intrinsic interaction
such as
\begin{equation}
{\cal S}=\displaystyle{\int{\rm
d}\tau\,\tr\sqrt{(D_{\tau}T)^{2}-(D_{\tau}X)^{2}\,}}\,.\label{NDBI}
\end{equation}
Quantization of the system is analogue to that of the string
theory. It would be interesting to see if     the critical
dimensions free of  the Lorentz symmetry anomaly exists. A single
particle system i.e. $1\times 1$ matrix model is  anomaly free in
any dimension. On the other hand,  the generic matrix model
including the off diagonal elements may be anomalous just like
the string theories where  the anomaly  comes from the
oscillators not from the zero mode.  Eq.(\ref{NDBI}) is a
straightforward generalization of the single relativistic
particle action, and an alternative approach to write the
relativistic matrix model would be  the dimensional reduction of
the non-Abelian Born-Infeld action  \cite{non-ABI}.

An electron in one place seems to be ``distinguishable" from one at far distance. To explain this large scale phenomenon,
 the spontaneous symmetry breaking of the $\mbox{U}(N)$. is to be considered.\\

{\bf Acknowledegement} The author  wishes to thank D. Bak, H.-W. Lee, M. Henneaux  and  K. Lee for valuable comments.


\end{document}